\documentclass[
preprint,
 amsmath,amssymb,
 aps,
showkeys,
showpacs
]{revtex4-2}

\usepackage{comment}
\usepackage{graphicx}
\usepackage{dcolumn}
\usepackage{bm}

\def\beq{\begin{equation}}
\def\eeq{\end{equation}}
\def\bea{\begin{eqnarray}}
\def\eea{\end{eqnarray}}

\begin{document}

\preprint{MS-TP-21-38}

\title{What a direct neutrino mass measurement might teach us \\ about the dark sector}
\thanks{Talk given at the 20th Lomonosov Conference on Element.\ Particle Physics, Moscow, August 19-25, 2021.
}%

\author{Michael Klasen}
 \email{michael.klasen@uni-muenster.de}
\affiliation{Institut f\"ur Theoretische Physik, Westf\"alische Wilhelms-Universit\"at M\"unster, Wilhelm-Klemm-Stra\ss{}e 9, 48149 M\"unster, Germany}

\date{\today}

\begin{abstract}
  Searches for Dark Matter suggest that it couples to ordinary matter only very
  weakly and possibly only through the Higgs or other scalar bosons. On the other hand,
  neutrinos might not couple to the Higgs boson directly, but only through a loop of Dark
  Matter particles, which would naturally explain the small neutrino masses. We demonstrate
  that current experimental constraints on such a ``scotogenic'' scenario allow to make the
  linear dependence of the lightest neutrino mass on the dark sector-Higgs coupling explicit,
  so that a measurement by the KATRIN experiment would directly determine its value.
\end{abstract}

\maketitle


\section{Introduction}
\label{sec:1}

The nature of Dark Matter (DM) and the absolute neutrino mass scale are two prominent
research topics, which are currently under intense scrutiny. In particular, we now have
observational evidence for DM ranging from galactic rotation curves to the large scale
structure of the Universe and the Cosmic Microwave Background (CMB), informing us that
DM is cold and about five times more abundant than ordinary matter. Neither Massive
Compact Halo Objects nor Primordial Black Holes or Standard Model (SM) neutrinos can
explain a substantial fraction of DM. New heavy particles such as Weakly Interacting
Massive Particles (WIMPs) remain the most promising candidates, since their relic
density after freeze-out agrees with observations \cite{Klasen:2015uma}.

\clearpage

We also know now from solar, atmospheric and reactor observations that (at least two)
SM neutrinos have non-zero masses with a minimal value for their sum of 0.06 eV for
normal mass ordering, which rises to 0.2 eV for each neutrino in the quasi-degenerate
regime \cite{ParticleDataGroup:2020ssz}. This happens to be also the sensitivity goal of the
KATRIN experiment \cite{Drexlin:2013lha}, which currently sets an upper limit of 0.8 eV on
the effective electron (anti-)neutrino mass \cite{Aker:2021gma}. Depending on theoretical
assumptions and the data sets included, cosmological upper limits on the sum of neutrino
masses can be as strong as 0.12 eV \cite{Planck:2018vyg}.

Scotogenic models provide an intriguing connection between the two puzzles of the nature
of DM and the smallness of neutrino masses. There, these neutrino masses are generated
at one or more loops, and at least one heavy particle in the loop can be a DM candidate.
A discrete $Z_2$ symmetry is usually employed to guarantee its stability and prevent a
tree-level seesaw mechanism.

\section{The scotogenic model}
\label{sec:2}

\begin{figure}[h!]
\includegraphics[width=0.5\textwidth]{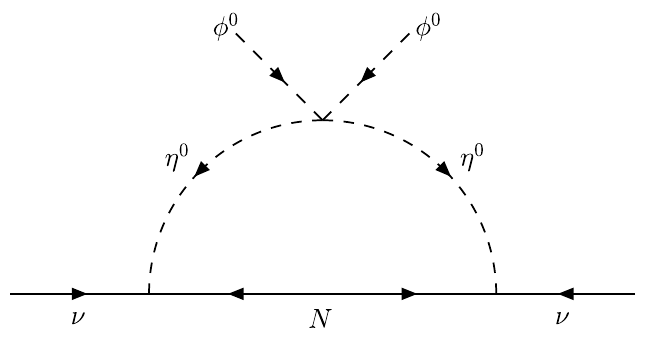}
\caption{\label{fig:01}Neutrino mass generation in the scotogenic model.}
\end{figure}

The simplest scotogenic (i.e.\ ``created from DM'') model is very economical,
as it adds to the left-handed SM lepton doublets $L_\alpha$ ($\alpha=1,2,3$)
only three generations of fermion singlets $N_i$ ($i=1,2,3$) and an inert
complex scalar doublet $(\eta^+,\eta^0)$, which does not develop
a vacuum expectation value (VEV). The lightest neutral fermion is then a good
DM candidate. Apart from kinetic terms, the Lagrangian of this model is given
by \cite{Ma:2006km}
\beq
 \mathcal{L}_N=-\frac{m_{N_i}}{2}N_iN_i+y_{i\alpha}(\eta^\dagger L_\alpha) N_i
+\mathrm{h.c.}-V
\eeq
with a scalar potential
\bea
V&=&
m_{\phi}^2\phi^\dag\phi+m_\eta^2\eta^\dag\eta
+\!\frac{\lambda_1}{2}\left(\phi^\dag\phi\right)^2
\!+\!\frac{\lambda_2}{2}\left(\eta^\dag\eta\right)^2
\!+\lambda_3\left(\phi^\dag\phi\right)\left(\eta^\dag\eta\right)\nonumber\\
&+&
\lambda_4\left(\phi^\dag\eta\right)\left(\eta^\dag\phi\right)
+\frac{\lambda_5}{2}\left[\left(\phi^\dag\eta\right)^2
+\left(\eta^\dag\phi\right)^2\right],
\eea
which breaks $SU(2)_L\times U(1)_Y\to U(1)_{\rm em.}$ when the SM Higgs $\phi$
obtains a VEV. Perturbativity imposes $|y_{i\alpha}|^2<4\pi$ and $|\lambda_{2,3,4,5}|<4\pi$,
and vacuum stability requires $\lambda_{1,2} > 0$, $\lambda_3 > - \sqrt{\lambda_1 \lambda_2}$
and $\lambda_3 + \lambda_4 - |\lambda_5|>- \sqrt{\lambda_1 \lambda_2}$.
Neutrino masses are then generated at one loop by the 3$\times$3
Yukawa matrices $y_{i\alpha}$ through the diagram in Fig.\ \ref{fig:01}.
A key observation is that the mass splitting of the real and imaginary components of
$\eta^0=(\eta_R+i\eta_I)/\sqrt{2}$, $m_R^2-m_I^2=2\lambda_5\langle\phi^0\rangle^2$, is
naturally small, since $\lambda_5=0$ implies $L$ conservation and $m_{\nu_i}=0$. In
the limit $\lambda_5\ll1$, the neutrino mass matrix simplifies to
\bea
(m_\nu)_{\alpha\beta}&\approx&2\lambda_5 \langle\phi^0\rangle^2
\sum_{i=1}^3\frac{y_{i\alpha} y_{i\beta}m_{N_i}}{32\pi^2(m_{R,I}^2-m_{N_i}^2)}
  \left[1+\frac{m_{N_i}^2}{m_{R,I}^2-m_{N_i}^2}
\log\left(\frac{m_{N_i}^2}{m_{R,I}^2}\right)\right],
\label{eq:3}
\eea
i.e.\ it is not only bilinear in $y$, but also linear in $\lambda_5$.

\section{Experimental constraints}
\label{sec:3}

Our main new observations are that the observational constraints on the DM relic density,
neutrino mass differences and mixing angles, lepton flavor violation (LFV) processes and new
charged particle masses allow us to make the linear dependence of the lightest neutrino
mass on the dark sector coupling $\lambda_5$ and the quadratic dependence on the eigenvalues
of the Yukawa couplings $y$ explicit \cite{deBoer:2020yyw}. To this end,
we impose the most recent measurements and limits from Planck \cite{Planck:2018vyg},
solar, atmospheric and reactor neutrinos \cite{ParticleDataGroup:2020ssz}, MEG
\cite{MEG:2016leq}, SINDRUM \cite{SINDRUM:1987nra} and OPAL \cite{OPAL:2003zpa}
on the parameter space of the scotogenic model, using the Casas-Ibarra parametrisation.
Direct and indirect detection constraints are not relevant, as the
fermion scatters off nuclei only at one loop. The lighest neutrino mass is scanned from
4 meV to 2 eV and $|\lambda_5|$ from $10^{-12}$ to $10^{-8}$.

\section{Numerical results}
\label{sec:4}

\begin{figure}[h!]
\includegraphics[width=0.7\textwidth]{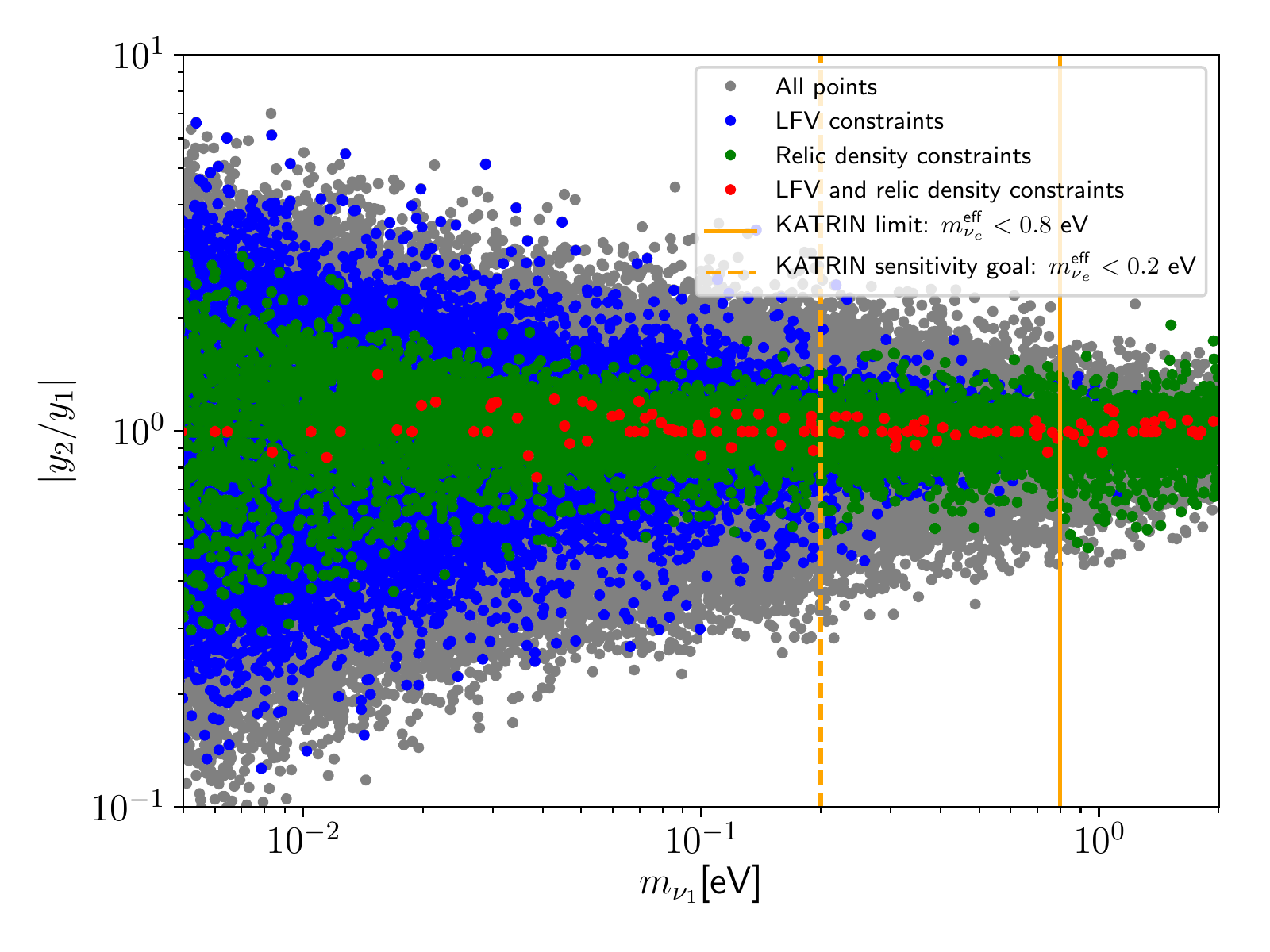}
\caption{\label{fig:02}Ratio of Yukawa couplings as a function of the lightest neutrino mass $m_{\nu_1}$ with mass difference/mixing (grey), LFV (blue), relic density (green) and all constraints (red points).}
\end{figure}

The experimental information mentioned above constrains the eigenvalues of the Yukawa
coupling matrices to be of similar size (Fig.\ \ref{fig:02}). For large absolute
neutrino masses, their differences and thus also those of the Yukawa couplings become
naturally small, whereas they are of course substantial in the limit of small absolute
neutrino masses (grey points). The upper limits on LFV processes (blue points) impose
upper limits on the Yukawas, while the relic density (green points) imposes lower
limits. Together, both constraints then lead to a narrow band of $|y_2/y_1|\sim1$
(red points) for both normal ordering (NO, shown here), inverse ordering (IO) and all
other combinations of Yukawa couplings (not shown).

\begin{figure}[h!]
\includegraphics[width=0.7\textwidth]{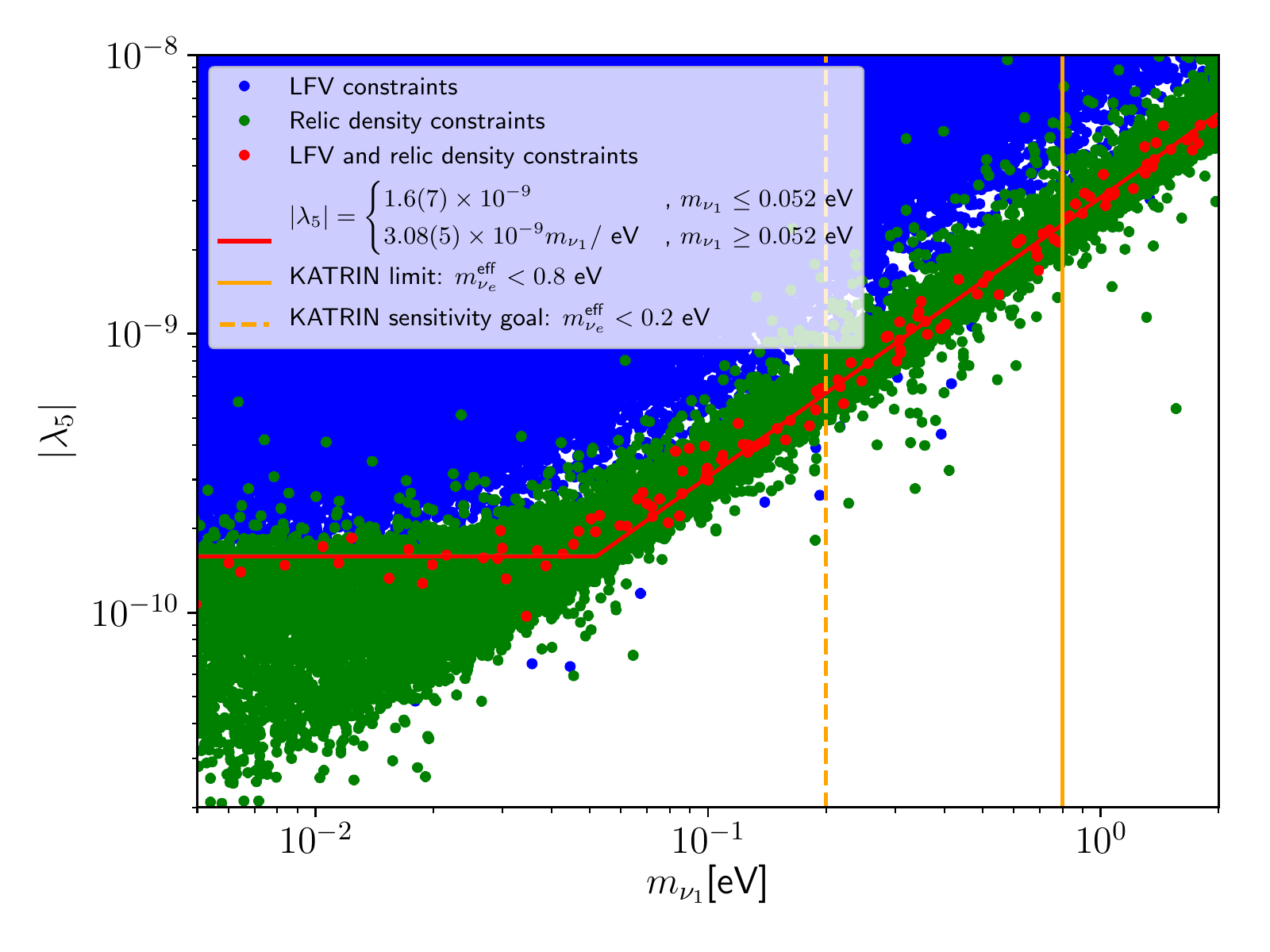}
\caption{\label{fig:03}The dark sector-Higgs boson coupling $|\lambda_5|$ as a function of lightest neutrino mass $m_{\nu_1}$ with LFV (blue), relic density (green) and all constraints (red points).}
\end{figure}

The linear dependence of the absolute neutrino mass on the dark sector-Higgs coupling
then becomes explicit (Fig.\ \ref{fig:03}), since LFV and relic density constraints act
again in opposite directions. A KATRIN measurement of the absolute neutrino mass would
therefore directly translate into a measurement of
\beq
|\lambda_5|=\left\{\begin{array}{ll}
(3.08\pm0.05)\times10^{-9} \ m_{\nu_1}/{\rm eV} & ({\rm NO})\\
(3.11\pm0.06)\times10^{-9} \ m_{\nu_1}/{\rm eV} & ({\rm IO})\end{array}\right. .
\eeq
Below $m_{\nu_1}=0.052$ eV, the heaviest neutrino mass dominates and
\beq
|\lambda_5|=\left\{\begin{array}{ll}
(1.6\pm0.7)\times10^{-10} & ({\rm NO})\\
(1.7\pm1.5)\times10^{-10} & ({\rm IO})\end{array}\right.\hspace*{19mm}
\eeq
becomes independent of $m_{\nu_1}$. The dark sector-Higgs boson coupling $\lambda_5$ can
therefore be predicted (modulo an arbitrary sign), once the absolute neutrino mass scale
is known.

\begin{figure}[h!]
\includegraphics[width=0.7\textwidth]{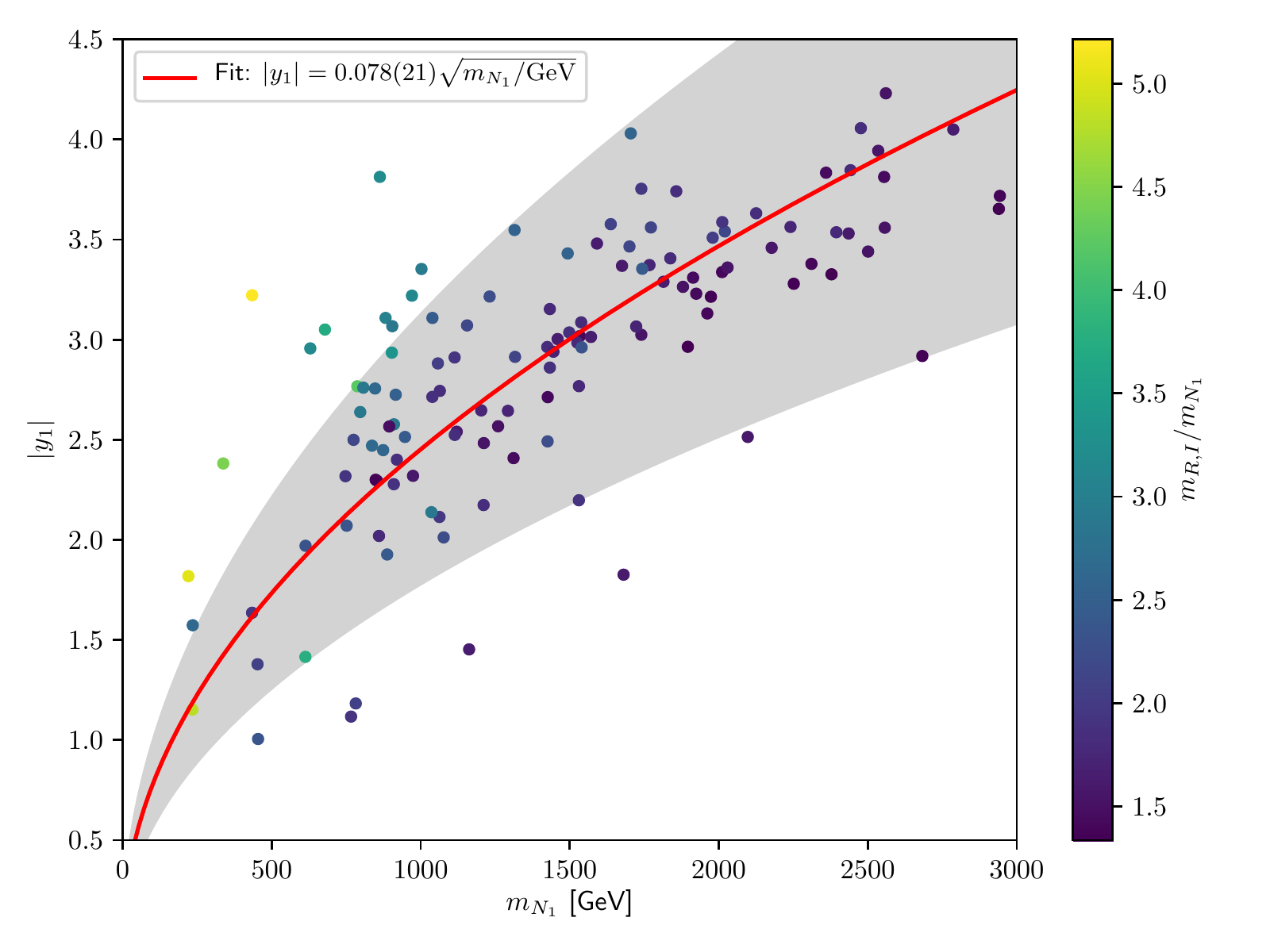}
\caption{\label{fig:04}Yukawa coupling of the lightest neutrino as a function of the DM mass. The ratio of the neutral scalar over the DM mass is given on the temperature scale.}
\end{figure}

With the ratio of $m_{\nu_1}/\lambda_5$ fixed, Eq.\ (\ref{eq:3}) can be inverted to
yield an approximate dependence
\beq
 |y_1|=\left\{\begin{array}{ll}
 (0.078\pm0.021) \ \sqrt{m_{N_1}/{\rm GeV}} & ({\rm NO})\\
 (0.081\pm0.012) \ \sqrt{m_{N_1}/{\rm GeV}} & ({\rm IO})\end{array}\right. . \nonumber
\eeq
of the lightest Yukawa coupling eigenvalue on the square root of the DM mass $m_{N_1}$
(Fig.\ \ref{fig:04}). The only condition is that the DM mass is sufficiently smaller
from those of the scalars (temperature scale) and much smaller than those of the other
fermions, which is imposed by the experimental constraints. This implies that if the DM
mass is known, we can predict its coupling to the SM charged leptons and neutrinos.

\begin{figure}
\includegraphics[width=0.66\textwidth]{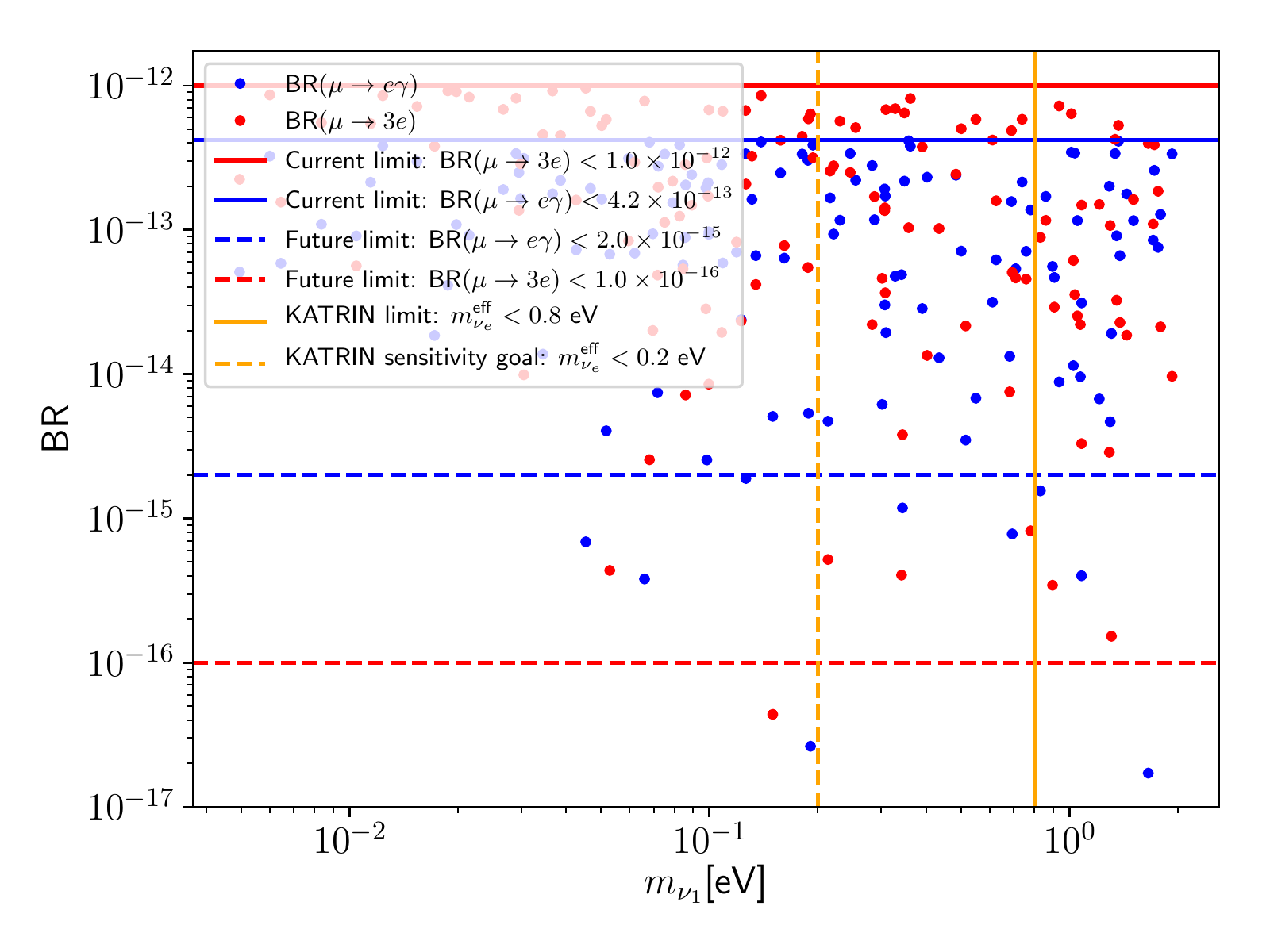}
\caption{\label{fig:05}Branching ratios of viable scotogenic models for the LFV processes $\mu\to e\gamma$ (blue) and $\mu\to3e$ (red points), their current (full) \cite{MEG:2016leq,SINDRUM:1987nra} and future (dashed) \cite{Renga:2018fpd,Blondel:2013ia} experimental limits, and the current \cite{Aker:2021gma} and future \cite{Drexlin:2013lha} KATRIN limits (yellow lines) on the electron neutrino mass.}
\end{figure}

As LFV and relic density constraints are not only complementary to each other, but also
to a direct neutrino mass measurement, the parameter space of fermion DM in the scotogenic
model will be almost completely testable in the near future (Fig.\ \ref{fig:05}).
Here, only points satisfying the neutrino mass difference, mixing angle and the
DM relic density constraints are shown. Currently the limit on $\mu\to e\gamma$
\cite{MEG:2016leq} is currently stronger than the one for $\mu\to3e$
\cite{SINDRUM:1987nra}, but this might change soon \cite{Renga:2018fpd,Blondel:2013ia}.

\section{Discussion and outlook}
\label{sec:5}

The theoretical reason for our observations lies in the intimate topological connection
of the neutrino mass diagram to the diagrams mediating LFV and, after cutting the internal
fermion line, DM annihilation. These correlations are absent for scalar DM, which can
also annihilate into weak gauge bosons. In this case, inelastic scattering in the Sun
can provide stringent bounds \cite{deBoer:2021pon}. If scalars and fermions are close
in mass, coannihilations must be and have been considered \cite{Klasen:2013jpa}.

\begin{acknowledgments}
  The author thanks T.\ de Boer, C.\ Rodenbeck and S.\ Zeinstra for their collaboration
  and the organisers of the conference for the invitation.
  This work has been supported by the DFG through the Research Training Network 2149
  ``Strong and weak interactions - from hadrons to dark matter''.
\end{acknowledgments}

\clearpage

%


\end{document}